\documentclass[prl,twocolumn,showpacs,floatfix,amsfonts]{revtex4}
\usepackage{graphicx,graphics,color,epsfig}
\usepackage{bm}
\usepackage{amsmath}
\usepackage{amssymb}
\usepackage{epstopdf}

\newcommand{\beq}{\begin{equation}}
\newcommand{\eeq}{\end{equation}}
\newcommand{\beqa}{\begin{eqnarray}}
\newcommand{\eeqa}{\end{eqnarray}}

\begin{document}

\title{Kondo Stripes in an Anderson-Heisenberg Model of Heavy Fermion Systems}

\author{Jian-Xin Zhu}
\email{jxzhu@lanl.gov}
\homepage{http://theory.lanl.gov}
\affiliation{Theoretical Division, Los Alamos National Laboratory, Los Alamos,
New Mexico 87545, USA}
\author{I. Martin}
\affiliation{Theoretical Division, Los Alamos National Laboratory, Los Alamos,
New Mexico 87545, USA}
\author{A. R. Bishop}
\affiliation{Theoretical Division, Los Alamos National Laboratory, Los Alamos,
New Mexico 87545, USA}

\date{\today}
\begin{abstract}
We study the interplay between the spin-liquid and Kondo physics, as related to
the non-magnetic part of the phase diagram of heavy fermion materials.  Within
the unrestricted mean-field treatment of the infinite-$U$ 2D
Anderson-Heisenberg model, we find that there are two topologically distinct
non-degenerate uniform heavy Fermi liquid states that may form as a consequence
of the Kondo coupling between spinons and conduction electrons.  For certain
carrier concentrations the uniform Fermi liquid becomes unstable with respect to formation of a new kind of anharmonic ``Kondo stripe" state with inhomogeneous Kondo screening strength and the charge density modulation.  These feature are experimentally measurable, and thus may help to establish the relevance of the spin-liquid correlations to heavy fermion materials.  

\end{abstract}
\maketitle

The intermetallic heavy fermion compounds based on either rare earth elements
or on actinides are prototypical strongly correlated
systems~\cite{GRStewart84,HTsunetsugu97,GRStewart01}. In these materials, there
are two types of electrons: delocalized conduction electrons ($c$-electrons),
which derive from the outer atomic orbitals, and strongly localized
$f$-electrons that singly occupy the inner orbitals. The interplay between the
$c$-$f$ hybridization and the screened on-site Coulomb repulsion, responsible
for the single occupancy of $f$ orbitals, gives rise to wide range of
behaviors, including magnetic ordering, Fermi liquid and non-Fermi
liquid~\cite{PGegenwart07}.  The magnetism arises primarily through the
spin-ordering in the localized $f$-band.  The heavy Fermi liquid (HFL) in the
non-magnetic phase is believed to occur via Kondo hybridization of the $f$ and
$c$ bands.  The non-Fermi liquid behavior emerges in the vicinity of the
quantum critical point separating these phases.

Recently, Senthil {\em et al.}~\cite{senthilPRL,senthilPRB} proposed a new route to a HFL state.
They considered the possibility that due to the frustrated magnetic interaction between spins in the $f$ band, instead of forming an ordered magnetic state at
low temperatures, a magnetically featureless {\em spin liquid} emerges
with spin-1/2 chargeless fermionic excitations (spinons). Upon inclusion
of strong enough Kondo interaction between the spinons and conduction
electrons, a HFL would form, with spinons acquiring an electrical charge~\cite{col_mar_scho} and merging with conduction electrons to form a ``large" Fermi surface.  An important distinction between
this scenario and the more conventional one~\cite{Multi2}, which does not include the
possibility of a spin liquid, is that the transition occurs at a
finite value of the Kondo coupling.  This results in a new kind of scaling
behavior near the quantum critical point separating the FL$^*$ phase of two
decoupled -- electron and spinon -- Fermi liquids and the HFL phase of coupled
spinons and electrons~\cite{senthilPRB,col_mar_scho,CPepin07}. Whether such a
quantum phase transition is realized in nature, or always preempted by a
conventional Landau phase transition~\cite{Multi1}, remains to be seen.

Spin liquid phases have proven to be very elusive as the ground states of real
materials, as well as microscopic models, particularly ones involving real
electrons. However, at finite temperatures, in systems with frustrated magnetic
interactions, magnetically disordered phases are well approximated by spin
liquids, even if at very low temperatures a small magnetic order parameter
emerges.  Therefore, it is natural to ask if the spin liquid tendencies can
manifest,  if not in the zero-temperature phase transition itself, in the
structure of the ``ordered" state away from the phase transition.  In this
Letter, we analyze in detail the structure of the heavy Fermi liquid that
emerges out of the coupling of the spin liquid and the conduction Fermi liquid
in the Anderson-Heisenberg lattice model.  Our main findings are: First, that
starting from the same spin liquid state one can obtain more than one kind of 
uniform heavy Fermi liquid. Second, that the spinon-electron Kondo instability
can be generically inhomogeneous. The resulting structure has inhomogeneous
distributions of Kondo couplings, as well as anharmonic charge density
modulation, very similar to the ``stripes'' in the Hubbard model~\cite{zaanen}. A
harmonic analog of this phase (without charge modulation) was recently found in
a continuum model by Paul {\em et al.}~\cite{IPaul07}.

We start from the Anderson-Heisenberg~\cite{ACHewson93} lattice model with
localized $f$-band,
\begin{eqnarray}
H &=& - \sum_{ij,\sigma} (t_{ij}^{c}  + \mu\delta_{ij})c_{i\sigma}^{\dagger}
c_{j\sigma} + \sum_{ij,\sigma} [V_{ij} c_{i\sigma}^{\dagger}\tilde{f}_{j\sigma}
+ \text{h.c.}]
\nonumber \\
&&  + \sum_{i\sigma}  (\epsilon_{f} - \mu)\tilde{f}_{i\sigma}^{\dagger}
\tilde{f}_{i\sigma} + \sum_{i} Un_{i\uparrow}^{f} n_{i\downarrow}^{f} \nonumber \\
&& + \frac{J_{H}}{2} \sum_{ij} \biggl{(}  \mathbf{S}_{i} \cdot \mathbf{S}_{j}
- \frac{\tilde{n}_{i}^{f} \tilde{n}_{j}^{f}}{4} \biggr{)}\;.
\label{EQ:Hamil1}
\end{eqnarray}
Here $c_{i\sigma}$ ($c_{i\sigma}^{\dagger}$) annihilates (creates) a conduction electron with spin $\sigma$ on site $i$;  $\tilde{f}_{i\sigma}$
($\tilde{f}_{i\sigma}^{\dagger}$) represents the $f$-electron. The number
operators for $c$ and $f$ orbitals with spin $\sigma$ are given by
$n_{i\sigma}^{c}=c_{i\sigma}^{\dagger}c_{i\sigma}$ and
$\tilde{n}_{i\sigma}^{f}=\tilde{f}_{i\sigma}^{\dagger}\tilde{f}_{i\sigma}$,
respectively. The quantity $t_{ij}^{c}$ is the hopping integral of the
conduction electrons, and  $\epsilon_{f}$ is the local energy level of the $f$
orbitals. The hybridization between the conduction and $f$ bands is represented by $V_{ij}$. The $f$-electrons experience the Coulomb repulsion of strength
$U$. We have included explicitly the superexchange (Heisenber) interaction between $f$-electron spins, $\mathbf{S}_{i} = \frac{1}{2} \sum_{\alpha\beta}
\tilde{f}_{i\alpha}^{\dagger} \bm{\sigma}_{\alpha\beta} \tilde{f}_{i\beta}$
with $\bm{\sigma}$ the Pauli matrix.  The chemical potential $\mu$
is introduced to tune the mismatch between the $c$ and $f$ patches of Fermi
surface.

For simplicity, we study here the $U\rightarrow \infty$ limit, which excludes
the double occupancy of the $f$-sites.  This limit explicitly breaks the
particle-hole symmetry in the $f$ band.  We find however, that results do not
qualitatively change for the Kondo-Heisenberg model~\cite{senthilPRB}, which
preserves the symmetry.  The constraint can be treated by means of the
``slave-particle'' representation~\cite{SBarnes76,PColeman84}
$\tilde{f}_{i\sigma} = f_{i\sigma} b^{\dagger}$, which replaces the original
fermion with two auxiliary particles -- a fermion $f$ (spinon) and a boson
$b$ (holon), satisfying the local constraint: $\sum_{\sigma}
f_{i\sigma}^{\dagger}f_{i\sigma} + b_{i}^{\dagger} b_{i} =1$. Further,
decoupling the superexchange interaction via the Hubbard-Stratonovich
transformation, the Hamiltonian ~(\ref{EQ:Hamil1}) can be written as
\begin{eqnarray}
H &=& - \sum_{ij,\sigma} (t_{ij}^{c}  + \mu\delta_{ij})c_{i\sigma}^{\dagger} c_{i\sigma}
+ \sum_{ij,\sigma} [V_{ij}b_{j}^{\dagger} c_{i\sigma}^{\dagger}f_{j\sigma} + \text{h.c.}]
\nonumber \\
&& - \sum_{ij,\sigma} [\chi_{ij} - (\epsilon_{f} + \lambda_{i}  - \mu )
\delta_{ij}]f_{i\sigma}^{\dagger} f_{j\sigma} \nonumber \\
&& + \sum_{i} \lambda_{i}(b_{i}^{\dagger}b_{i}-1)  + \sum_{ij}{\frac{|\chi_{ij}|^2}{J_H}}\;.
\label{EQ:Hamil2}
\end{eqnarray}
The single occupancy constraint on each site has been implemented through   a
Lagrange multiplier $\lambda_{i}$.  To proceed, we make a static approximation
by assuming that the slave bosons are frozen ($b_i\rightarrow \langle
b_i\rangle$) and the spin liquid parameters assume their mean field values,
$\chi_{ij} = (J_{H}/2) \langle f_{i\sigma}^{\dagger} f_{j\sigma} \rangle$.
Moreover, for the purpose of present discussion we assume that the uniform
spin liquid ($\chi_{ij}=\chi$~\cite{Baskaran87}) is a good reference state for
the formation of the Kondo Fermi liquid, thereby making $\chi$ an input
parameter of our mean field model. Note that there is a local $U(1)$ gauge
freedom present in this description: A change of phase of the slave boson on
site $i$, $b_i \rightarrow b_i e^{i\phi_i}$, with the simultaneous change $f_i
\rightarrow f_i e^{i\phi_i}$ and $\chi_{ij}\rightarrow e^{-i\phi_i}\chi_{ij}
e^{\phi_j}$ leaves the physical state unaltered.  There are alternative
mean-fields that are possible, including anomalous decoupling of the
superexchange term, similar to the $t$-$J$ model for cuprate
superconductors~\cite{lee}.  However, since our focus here is on possible
inhomogeneous Kondo phases, we defer this possibility to a future
consideration~\cite{JXZhu07}.  In the lattice space, the Bogliubov de-Gennes (BdG) mean-field equations are:
\begin{equation}
\sum_{j} \left(
\begin{array}{cc}
h_{ij}^{f} &  \Delta_{ij}   \\
\Delta_{ji}^{*}    & h_{ij}^{f}
\end{array} \right)
\left( \begin{array}{c}
u_{j\sigma}^{n} \\
v_{j\sigma}^{n}
\end{array} \right)  = E_{n}
\left( \begin{array}{c}
u_{i\sigma}^{n} \\
v_{i\sigma}^{n}
\end{array} \right) \;,
\label{EQ:BdG}
\end{equation}
subject to the constraints
\begin{equation}
\sum_{\sigma} \langle f_{i\sigma}^{\dagger} f_{i\sigma}\rangle + \vert b_{i} \vert^{2} = 1\;,
\end{equation}
and
\begin{equation}
\lambda_{i} b_{i} + \sum_{j\sigma} V_{ji} \langle c_{j\sigma}^{\dagger} f_{i\sigma} \rangle
-\sum_{j\sigma} t_{ji}^{f} \langle f_{j\sigma}^{\dagger} f_{i\sigma} \rangle = 0 \;.
\end{equation}
Here $h_{ij}^{c} = -t_{ij}^{c} -\mu \delta_{ij}$, $\Delta_{ij} = V_{ij}
b_{j}^{*}$, and $h_{ij}^{f} = -\chi_{ij} + (\epsilon_{f} +
\lambda_{i} -\mu)\delta_{ij} $.   
We solve this set of equations self-consistently via
exact diagonalization on a two-dimensional square lattice. For simplicity, we
consider on-site hybridization only, $V_{ij} = V\delta_{ij}$.  We note,
that a finite-ranged hybridization can lead to non-zero orbital momentum Kondo
states with the hybridization gap anisotropic in momentum
space~\cite{PGhaemi07}. The momentum-space inhomogeneity is complementary to
the real space one discussed here. Throughout this work, the quasiparticle
energy is measured with respect to the Fermi energy and the energy unit $t^c=1$
is chosen.

\begin{figure}[ht]
\includegraphics[width=.7\columnwidth]{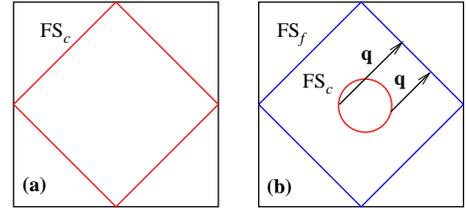}
\caption[]{(Color) A cartoon of possible Fermi surface topologies of conduction
electrons and spinons. (a) The effective mass of spinons is infinite ($\chi =
0$) and no spinon Fermi surface is present; (b) $\chi \ne 0$:  The Fermi
surfaces of the conduction electrons and spinons, in general, are mismatched.
 \label{FIG:FS}}
\end{figure}

 \begin{figure}[t]
\centerline{\psfig{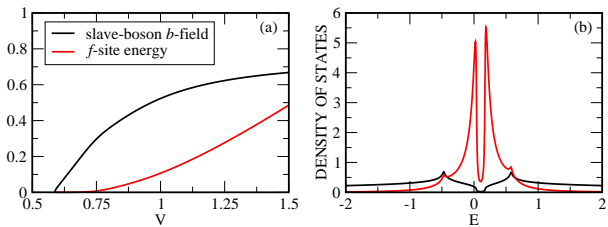}}
\caption{(Color) Dispersionless spinons, $\chi = 0$.  (a) The slave boson
$b$-field and the effective $f$-site energy $\epsilon_f + \lambda -\mu$, as a
function of $V$ for $\chi_{0} = \mu =0$.  (b) The projected conduction (black
line) and f-electron (red line) density of states,
 as a function of energy for $V = 1.0$. The temperature $T=0.01$. } \label{FIG:Boson1}
\end{figure}

Within the mean-field treatment outlined above, there are two distinct uniform
heavy Fermi liquid states.  Suppose that we fix the gauge such that the
spin-liquid parameter is positive, $\chi > 0$, and solve for Kondo
hybridization, $b_i$. One possible solution is a constant, $b_i = b_0$. Another
is $b_i =b_0 (-1)^{i_x + i_y}$, which naively appears to break translational
invariance. However, upon a gauge transformation it can be transformed to a
uniform state with $b_i =b_0$ and $\chi < 0$, which is clearly uniform.  The
two states can be distinguished based on ${\rm sign}(b_i\chi_{ij}b_j)$
\cite{sachdev}.  We will denote the Kondo phase with $b_i\chi_{ij}b_j>0$ as
``even" (eHFL) and $b_i\chi_{ij}b_j<0$ as ``odd" (oHFL).  For these two states,
we can solve the above set of equations in momentum space. For eHFL, the
energy dispersion has two branches given by:
\begin{equation}
E_{\mathbf{k},\pm} = Z_{\mathbf{k}} \pm \sqrt{Q_{\mathbf{k}}^{2} + \Delta_{\mathbf{k}}^{2}}\;,
\label{EQ:Energy_Uni}
\end{equation}
where $Z_{\mathbf{k}} (Q_{\mathbf{k}})=(\xi_{\mathbf{k}}^{c} \pm
\xi_{\mathbf{k}}^{f})/2$  with $\xi_{\mathbf{k}}^{c}= -2t^{c} (\cos k_x + \cos
k_y) -\mu$ and $\xi_{\mathbf{k}}^{f} = -2\chi(\cos k_x + \cos k_y) + (\lambda
-\mu + \epsilon_f)$, and $\Delta_{\mathbf{k}}=Vb_{0}$. For oHLF, one only needs
to replace $\chi \rightarrow -\chi$.

In Fig.~\ref{FIG:FS}, we provide a cartoon of possible Fermi topologies of
decoupled conduction electrons and spinons. When the spinon energy is
dispersionless, the Fermi surface is undefined, i.e. all spinons have the same
-- zero -- energy in the Brillouin zone, Fig.~\ref{FIG:FS}(a). This is the
situation of the standard Anderson lattice model.  When the spinons disperse in
the presence of a uniform spin liquid, the spinon Fermi surface is formed,
Fig.~\ref{FIG:FS}(b). Again, due to the single occupancy constraint, it is
pinned at half-filling, regardless of the chemical potential.  The Fermi
surface of the conduction electrons, FS$_c$, can be tuned by the chemical
potential in both cases.

First, as a point of reference, we consider the case of dispersionless spinons, $\chi = 0$.  In Fig.~\ref{FIG:Boson1}(a) we show the slave boson field (effective $c$-$f$ band hybridization) and the mean-field $f$-site energy, as a function of the coupling parameter $V$.  In Fig.~\ref{FIG:Boson1}(b), we show the projected
density of states (DOS).  There is a pronounced hybridization gap that appears
above the Fermi energy in both projected conduction and $f$-electron DOS.

\begin{figure}[b]
\centerline{\psfig{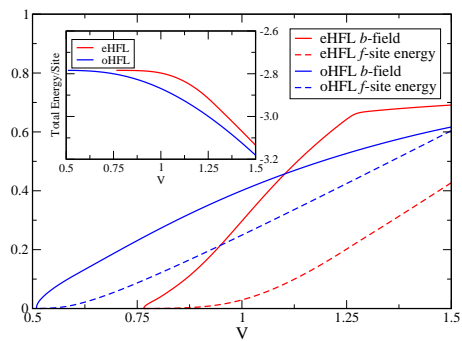}} 
\caption{(Color) The
slave boson $b$-field and the effective $f$-site energy $\epsilon_f + \lambda
-\mu$ for both the even and odd Kondo phases,  as a function
of $V$ for $\mu =0$ but $\chi=0.1$. Inset: The corresponding
Helmholtz free energy.
The temperature $T=0.01$. In this case, the odd Kondo state has a lower
energy than the even Kondo state. } \label{FIG:Boson2}
\end{figure}

\begin{figure}[t]
\centerline{\psfig{figure=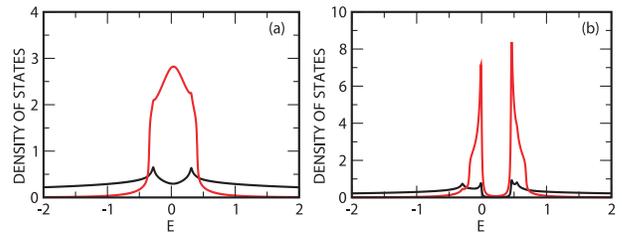,width=8cm,angle=0}} \caption{(Color) The
projected conduction (black lines) and $f$-electron (red lines) densities of states for
the even (a) and odd Kondo (b) states, corresponding to
Fig.~\ref{FIG:Boson2}. } \label{FIG:Boson2_Dos}
\end{figure}

Now we turn to the case of dispersive spinons.  For zero chemical potential
$\mu=0$, the uncoupled Fermi surfaces of the conduction electrons, FS$_c$, and
spinons, FS$_f$, coincide, i.e., the red line in Fig.~\ref{FIG:FS}(b) overlaps
the blue one.  In this case, there is a perfect nesting across the two Fermi
surfaces.  This has important consequences for the Kondo hybridization,
breaking the degeneracy between the odd and even Kondo phases, which existed in the $\chi = 0$ case.  Qualitatively, the energetic preference of the odd Kondo phase can be seen from the approximate correspondence between the present model and the repulsive Hubbard model.  Spin up (down) electrons in the Hubbard model correspond to $c$ ($f$) electrons here.  In the Hubbard model, the dominant instability at half filling occurs at momentum $(\pi,\pi)$ and corresponds to antiferromagnetism.  In the present model, antiferromagnetism clearly corresponds to the odd Kondo phase.

In Fig.~\ref{FIG:Boson2}, we present the self-consistent numerical results for the odd and even 
uniform Kondo phases.  As one can see from Fig.~\ref{FIG:Boson2}, the quantum critical points separating the FL$^*$ phase~\cite{senthilPRB} from the odd and even Kondo (heavy Fermi liquid) states occur at different values of interband coupling paramter $V$.  Comparison of free energies (see the inset of Fig.~\ref{FIG:Boson2}), as expected, reveals that the odd Kondo phase has lower energy than the even phase.  That the odd Kondo phase takes better advantage of the Fermi surface geometry is prominently manifested in the DOS.  For the even Kondo state, although there is a depression exhibited in the projected conduction electron DOS, the projected $f$-electron DOS shows a broad band behavior (see Fig.~\ref{FIG:Boson2_Dos}(a)). However, for the odd Kondo state, an $s$-wave-like gap opens in both the projected conduction electron and $f$-electron DOS (see  Fig.~\ref{FIG:Boson2_Dos}(b)).

\begin{figure}[t]
\includegraphics[width=1.0\columnwidth]{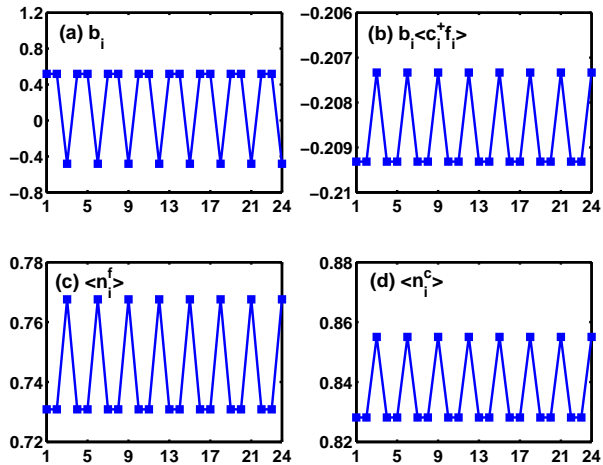}
\caption{(Color) Spatial variation along the $x$-direction of the slave-boson $b$ field (a),
the effective hybridization $b_{i}\langle c_{i}^{\dagger}f_{i}\rangle$ (b),  the averaged $f$-electron (c) and conduction electron (d) densities  for $V=0.8$ and $\mu=-0.50$. The
temperature $T=0.01$. The calculations are carried out on a lattice of
$24\times 24$ sites.} \label{FIG:KondoStripe}
\end{figure}

A finite chemical potential breaks the particle-hole symmetry in the conduction band and introduces a mismatch between the Fermi surfaces of the conduction electrons, FS$_c$,  and spinons, FS$_f$.  In this case, an instability to a state modulated at an incommensurate wavevector, related to the mismatch between FS$_c$ and FS$_f$, is possible. Indeed, in the continuum, an incommensurate {\em harmonic} Kondo wave has been recently found for circular Fermi
surfaces~\cite{IPaul07}.  Here, we explore the real-space modulation pattern, by solving the BdG equations~(\ref{EQ:BdG}) on a finite size ($24 \times 24$) lattice self-consistently via exact diagonalization.  To facilitate numerical convergence, we considered only unidirectional patterns.  This was achieved by ``seeding" them as an initial guess, and iterating until convergence.  We found that this procedure leads to convergence to a local minimum, without changing the initially imposed spatial period.  The optimal modulation pattern is determined by comparing the free energies of the converged solutions with the real-space modulation periods $N_x/n_x$ ($n_x$ is an integer running from 0 to $N_x-1$ and $N_x=24$ in our simulation).  In general, the spatial modulation pattern is {\em anharmonic}, in close analogy to the stripes found in the mean-field treatment of the Hubbard model \cite{zaanen}.  For a fixed coupling parameter $V$, we find that there is a critical value of $\mu$, below which a modulated Kondo state occurs.

In Fig.~\ref{FIG:KondoStripe} we show the spatial dependence of the slave-boson $b$ field, the Kondo hybridization $b_i\langle  c_i^\dagger  f_i\rangle$, the local $f$-electron density, and the local $c$-electron density (the latter three quantities are gauge-invariant).  We see clearly the emergence of an anharmonic {\em Kondo stripe} state  when the system  is deeply in the heavy fermion liquid phase.  In this strong coupling limit, the hole density in the  $f$-band is rather high, which leads to a short periodicity.  We also notice that in the strong coupling limit, the ``domain wall" between the regions of constant sign of $b_i$ is bond centered. We anticipate that by reducing the coupling strength $V$ toward the Kondo breakdown critical point, the anharmonicity will be suppressed along with a reduction of the modulation amplitude, leading to a nearly-harmonic Kondo wave~\cite{IPaul07}.  However, system size limitations and problems with numerical convergence in this parameter regime, prevented us from verifying this directly.
The Kondo stripes are closely analogous to the antiferromagnetic (AF) stripes in the mean-field treatment of the repulsive Hubbard model~\cite{zaanen}.  The field $b(\mathbf{r})$ plays the role of the AF order parameter $M(\mathbf{r})$, and the electron density $n(\mathbf{r})$ corresponds to the hole density 
$\rho(\mathbf{r})$.  The lowest-order symmetry-allowed coupling in the Kondo stripe free energy (which also follows from the constraint) is  $|b(\mathbf{r})|^2 n(\mathbf{r})$, which corresponds to 
$S^2(\mathbf{r})\rho(\mathbf{r})$ term for AF stripes~\cite{pryadko}.

In summary, we have considered a lattice model of coupled conduction electrons and spinons, as a model of a heavy Fermi liquid (Kondo) state.  We found that for nearest-neighbor hopping models, two distinct heavy Fermi liquid states are possible.  We also point out that in the generic case of mismatched electron and spinon Fermi surfaces, electronically inhomogneneous heavy Fermi liquid phases (Kondo stripes) are expected to appear.  These phases exhibit inhomogeneous Kondo hybridization and charge density, which should manifest in both local and bulk probes.  Identifying these phases
experimentally may help to determine the relevance of the spin-liquid physics to heavy fermion materials.

{\bf Acknowledgments:} We acknowledge useful discussions with C. D. Batista, C. P\'{e}pin, S. Sachdev and J. Zaanen. This work was carried out under the auspices of the National
Nuclear Security Administration of the U.S. Department of Energy at Los Alamos
National Laboratory under Contract No. DE-AC52-06NA25396 and supported by the
LANL/LDRD Program.

\end{document}